\documentclass[review]{elsarticle}

\usepackage{lineno,hyperref}
\modulolinenumbers[200]

\usepackage{amsmath}
\usepackage{graphicx}
\usepackage{xcolor}
\usepackage{epstopdf}
\usepackage[caption=false]{subfig}

\journal{Physics Letters A}

\begin{document}

\begin{frontmatter}

\title{Extensions of the General Solution Families for the Inverse Problem of the Calculus of Variations for Sixth- and Eighth-order Ordinary Differential Equations}


\author[mymainaddress]{S. Roy Choudhury \corref{mycorrespondingauthor}}
\cortext[mycorrespondingauthor]{Corresponding author}
\ead{sudipto.choudhury@ucf.edu}

\author[mymainaddress]{Ranses Alfonso-Rodriguez}

\address[mymainaddress]{Department of Mathematics, University of Central Florida, Orlando, FL, USA}

\begin{abstract}
New third- and fourth-order Lagrangian hierarchies  are derived in this paper. The free coefficients in the leading terms satisfy the most general differential geometric criteria currently known for the existence of a variational formulation, as derived by solution of the full inverse problem of the Calculus of Variations for scalar sixth- and eighth-order ordinary differential equations (ODEs). The Lagrangians obtained here have greater freedom since they require conditions only on individual coefficients. In particular, they contain four arbitrary functions, so that some investigations based on the existing general criteria for a variational representation are particular cases of our families of models. The variational equations resulting from our generalized Lagrangians may also represent traveling waves of various nonlinear evolution equations, some of which recover known physical models. For a typical member of our generalized variational ODEs, families of regular and embedded solitary waves are also derived in appropriate parameter regimes. As usual, the embedded solitons are found to occur only on isolated curves in the part of parameter space where they exist.

\end{abstract}


\end{frontmatter}

\linenumbers

\section{Introduction}
\label{sec:Intro}

There has been renewed interest in the derivations and use of Lagrangians for higher-order differential equations recently. Recent applications include, but are not limited to, higher-order field-theoretic models \cite{PUM, Man}, investigations of isochronous behaviors in a variety of nonlinear models \cite{TCG}, treatments of higher-order Painlev\'e equations \cite{CGK}, and variational treatments of embedded solitary waves of a variety of nonlinear wave equations \cite{VK}.

The inverse problem of the Calculus of Variations involves finding a Lagrangian for a given ordinary differential equation (ODE). The classical problem, for Lagrangians restricted to functions of the first derivatives of the dependent variable $u$ on $z$, has the form $u'' = F(z, u, u')$. The necessary and sufficient condition for such an equation to be derivable from the variational Euler-Lagrange equation%
\begin{align} \label{EL}
\partial _{u}L& -  d(\partial _{u'}L)/dz = 0,
\end{align}%
\noindent
are classical, and were first derived by Helmholtz \cite{Helm,Lop}, and also investigated by Darboux \cite{Dar}.

Juras \cite{Jur} has treated the general solution to the inverse problem for sixth- and eighth-order ODEs. We shall consider his main results later, and compare it to those derived in this paper.

This paper significantly extends the approach in \cite{Ran}, henceforth referenced as paper I, to sixth- and eighth-order variational ODEs. Treating the coefficients of the highest derivative in the ODE as arbitrary functions of the dependent variable and its derivatives generates families of Lagrangians for entire classes of differential equations. The resulting ODE coefficients satisfy the most general conditions currently known\cite{Jur} for variational sixth- and eighth-order ODEs. The variational ODEs which result are more general in that they leave several individual coefficients in the Lagrangian free, as well as allowing the resulting variational ODE to have arbitrary leading-order coefficients. Different types of solutions of various member equations in the corresponding families of variational ODEs may be analyzed using the corresponding generalized Lagrangian families. Regular and embedded solitary wave solutions of one member equation of the sixth-order variational ODE family are also derived as an illustrative topical example.

The remainder of this paper is organized as follows. 
Section \ref{sec:2} extends the treatment of the inverse problem developed in I for fourth-order variational ODEs to  general classes of nonlinear sixth- and eighth-order ODEs. By matching the ODE to the Euler-Lagrange variational equations, the general form of Lagrangian for all possible variational ODEs of each class is derived, together with auxiliary conditions on the other ODE coefficients. Section \ref{sec:3} then employs the resulting Lagrangian to construct both regular and embedded solitary waves for a representative member of this most general variational ODE of this family. Section \ref{sec:4} discusses the families of Lagrangians derived here and the resulting variational ODE classes, and makes specific comparisons to the most general ones currently known\cite{Jur}. Section \ref{sec:conclusions} briefly reviews the results and conclusions of the paper.


\section{Generalized Classes of Lagrangians and their associated variational ODEs}
\label{sec:2}


\subsection{Third-order Lagrangians}

In order to generalize third-order Lagrangian systems in a manner analogous to the treatment of fourth-order variational ODEs (obtained from second-order Lagrangians) in I, consider the leading-order sixth derivative term of a variational ODE to be of the most general form which one may get in the sixth-order Euler-Lagrange equation of a Lagrangian of the form $L(u, u', u", u^{(3)})$. Hence,  consider
\begin{equation}\label{leading}
-c_1(u,u',u'')c_2(u^{(3)})u^{(6)},
\end{equation}
and we want to match it to the Euler--Lagrange equation
\begin{equation}
    \dfrac{\partial L}{\partial u} -\dfrac{\partial }{\partial z}\left[ \dfrac{\partial L}{\partial u'} \right] +\dfrac{\partial^2 }{\partial z^2}\left[ \dfrac{\partial L}{\partial u''} \right] -\dfrac{\partial^3 }{\partial z^3}\left[ \dfrac{\partial L}{\partial u^{(3)}} \right]=0,
\end{equation}
whose full expression we present in \ref{sec:AppA}, Equation \eqref{Eq:EL}.

 Note that the form \eqref{leading} assumed above is more general than that employed in \cite{Jur}, and this will lead to more general variational equations than derived in that paper.

Thus, at order $O\left(u^{(6)}\right):$ we get
\begin{equation}\label{Eq:Coeff2}
L_{u^{(3)}u^{(3)}}=c_1(u,u',u'')c_2(u^{(3)}),
\end{equation}
or, on integrating, 
$$ 
	L_{u^{(3)}}=c_1(u,u',u'')\int c_2(u^{(3)}) du^{(3)}+c_3(u,u',u'').
$$ 
A second integration now yields:
\begin{equation}\label{Eq:Coeff3}
L=c_1(u,u',u'')c_4(u^{(3)})+c_3(u,u',u'')u^{(3)}+c_5(u,u',u''),
\end{equation}
where 
\begin{equation}\label{Eq:IntCond}
c_4(u^{(3)})={\displaystyle \iint c_2(u^{(3)})du^{(3)}du^{(3)} }  .
\end{equation}

Hence, Equation \eqref{Eq:EL} defines the most general variational sixth order ODE (which could be the traveling wave equation of some nonlinear partial differential equation (PDE) of interest, as in \cite{VPh, Smi09}), and can now be written as
\begin{multline} \label{vareq}
	u^{(6)}Z_1+u^{(5)}E_1+{u^{(4)}}^3\Delta_3+{u^{(4)}}^2\Delta_2+{u^{(4)}}\Delta_1 +{u^{(3)}}^3\Gamma_3+{u^{(3)}}^2\Gamma_2+{u^{(3)}}\Gamma_1 \\
	+{u''}^3B_3+{u''}^2B_2+{u''}B_1 +{u'}^3A_3+{u'}^2A_2+{u'}A_1 +L_{u}=0, 
\end{multline}
where $Z_1$, $E_1$, $\Delta_1$, $\Delta_2$, $\Delta_3$, $\Gamma_1$, $\Gamma_2$, $\Gamma_3$, $B_1$, $B_2$, $B_3$, $A_1$, $A_2$ and $A_3$ are given in Appendix \ref{sec:AppA}.

In Section 4 we shall prove that \eqref{vareq} generalizes the most general variational sixth-order ODEs known currently, as derived by Juras\cite{Jur}, and which, in fact, turn out to be particular cases of \eqref{vareq}.

\subsection{Fourth-order Lagrangians}

To generalize fourth-order Lagrangian systems in analogous fashion, consider the eighth derivative term in a variational nonlinear ODE, corresponding to the Euler-Lagrange equation, to be of the very general form
\begin{equation} d_1(u,u',u'', u^{(3)})d_2(u^{(4)})u^{(8)}.
\end{equation} \label{fourth}
and we want to match it to the Euler--Lagrange equation
\begin{equation}
    \dfrac{\partial L}{\partial u} -\dfrac{\partial }{\partial z}\left[ \dfrac{\partial L}{\partial u'} \right] +\dfrac{\partial^2 }{\partial z^2}\left[ \dfrac{\partial L}{\partial u''} \right] -\dfrac{\partial^3 }{\partial z^3}\left[ \dfrac{\partial L}{\partial u^{(3)}} \right] + \dfrac{\partial^4 }{\partial z^4}\left[ \dfrac{\partial L}{\partial u^{(4)}} \right]=0.
\end{equation}

Thus, at order $O\left(u^{(8)}\right):$ we get
\begin{equation}\label{Eq:Coeff24}
L_{u^{(4)}u^{(4)}}=d_1(u,u',u'',u^{(3)})d_2(u^{(4)}),
\end{equation}
or, on integrating, 
$$ 
	L_{u^{(4)}}=d_1(u,u',u'',u^{(3)})\int d_2(u^{(4)}) du^{(4)}+d_3(u,u',u'',u^{(3)}).
$$ 
A second integration now yields:
\begin{equation}\label{Eq:Coeff34}
L=d_1(u,u',u'',u^{(3)})d_4(u^{(4)})+d_3(u,u',u'',u^{(3)})u^{(4)}+d_5(u,u',u'',u^{(3)}),
\end{equation}
where 
\begin{equation}\label{Eq:IntCond4}
d_4(u^{(4)})={\displaystyle \iint d_2(u^{(4)})du^{(4)}du^{(4)} }  .
\end{equation}

Hence, for this family of Lagrangians, the Euler-Lagrange equation above gives the variational eighth order ODE. This equation may now be expanded in terms of the $d_i$'s using equation \eqref{Eq:Coeff34}, and put into the form in equation \eqref{vareq} above. In Section 4 we prove that the resulting variational eighth-order ODE extends the most general families of variational eighth-order variational ODEs known currently, as derived by Juras\cite{Jur}, and which, in fact, turn out to be particular cases of this equation. However, we omit the explicit equation for the time being since it is lengthy, and may be directly obtained by computer algebra using the intermediate formulas and steps given above.

\subsection{Two Cases Worth Mentioning}
\label{subsec:3.1}

Notice that for the very general case 
$$ 
	c_1(u,u',u'')=\left( \sum_{k=0}^r a_k u^k \right)\left( \sum_{l=0}^s b_l (u')^l \right)\left( \sum_{l=0}^t e_m (u'')^m \right) ,
$$ 
one may directly obtain the variational ODE \eqref{vareq}. 

Clearly, an analogous general case for the fourth-order Lagrangians \eqref{Eq:IntCond4} would consider
$$ 
	d_1(u,u',u'',u^{(3)})=\left( \sum_{k=0}^r a_k u^k \right)\left( \sum_{l=0}^s b_l (u')^l \right)\left( \sum_{m=0}^t f_m (u'')^m \right) \left( \sum_{n=0}^u g_n (u^{(3)})^n \right).
$$ 

We consider a specific example of the first class in detail in the following section. Following that, we consider the general case and compare our results to those obtained earlier for the existence of Lagrangians of sixth- and eighth-order ODEs.


\section{A member of the family of generalized Lagrangians}
\label{sec:3}

If in the previous section we set 
\begin{align}
	c_1(u,u') & =-d_1+d_2u+d_3u'+d_4u'',&  c_2(u^{(3)}) & =1,\\  c_3(u,u') & =u^2, & c_5(u,u') & =u'^2,
\end{align}
then the Lagrangian reads
\begin{equation}
L=\left(-d_1+d_2u+d_3u_z+d_4u_{zz}\right)u_{zzz}^2+u^2 u_{zzz}+u_z^2,
\end{equation}
and its Euler--Lagrange equation is
\begin{multline}\label{odeg}
u''\left[ 6u'+2 \right]+\dfrac{d_2}{2}{u^{(3)}}^2+u^{(4)}\left[ 2d_4u^{(4)}+5d_3u^{(3)}+3d_2u'' \right] \\
 +3u^{(5)}\left[ d_4u^{(3)}+d_3u''+d_2u' \right] +u^{(6)}\left[ d_4u''+d_3u'+d_2u-d_1 \right]=0.
\end{multline}

In the next subsections we will apply the Rayleigh-Ritz method and its extensions\cite{Ran, VPh} to obtain solitary wave solutions for the variational ODEs of this general class of Lagrangians.

\subsection{Regular solitons}
\label{subsec:4.1}
Using the same trial function and following the process outlined in I, one gets the averaged action

\begin{equation}
S=\dfrac{\sqrt{\pi}A^2\left(32\sqrt{3}A\left( -16d_4+21d_2\rho^2 \right) +81\sqrt{2}\rho^2\left( -15d_1+2\rho^4 \right) \right)\sqrt{\pi}}{324\rho^7}.
\end{equation}

\noindent
By solving the system of equations generated by differentiating the action with respect to the parameters, one gets a non--trivial solution for $A$ and $\rho$ that allow us to evaluate the regular soliton for different values of the parameters. 

\begin{figure*}[ht]
   \subfloat[The regular soliton for various $d_1$ values.]{\label{Fig1:a}
      \includegraphics[width=.47\textwidth]{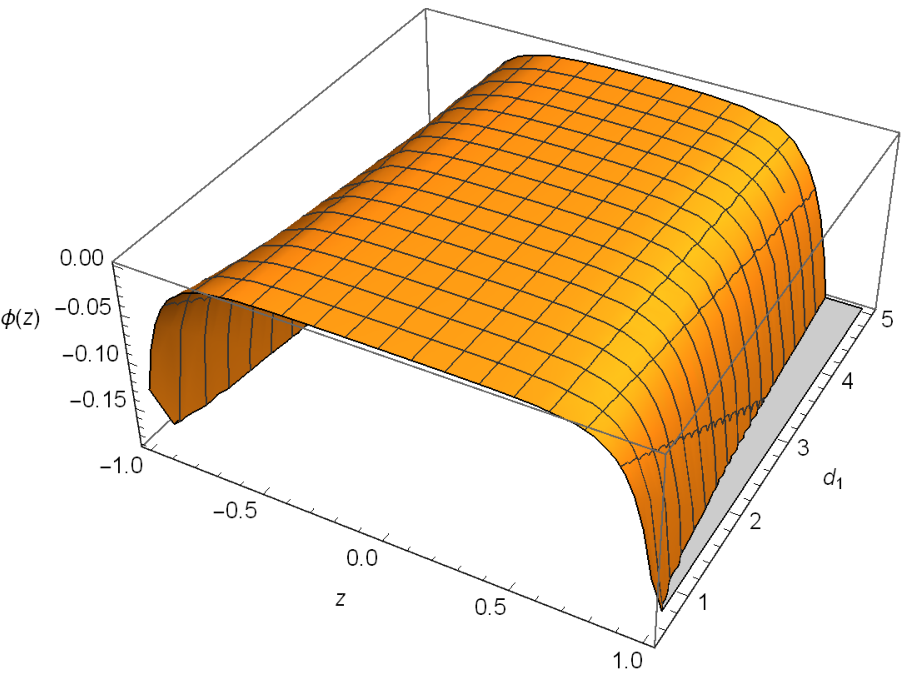}}
~
   \subfloat[The regular soliton for various $d_2$ values.]{\label{Fig1:b}
     \includegraphics[width=.47\textwidth]{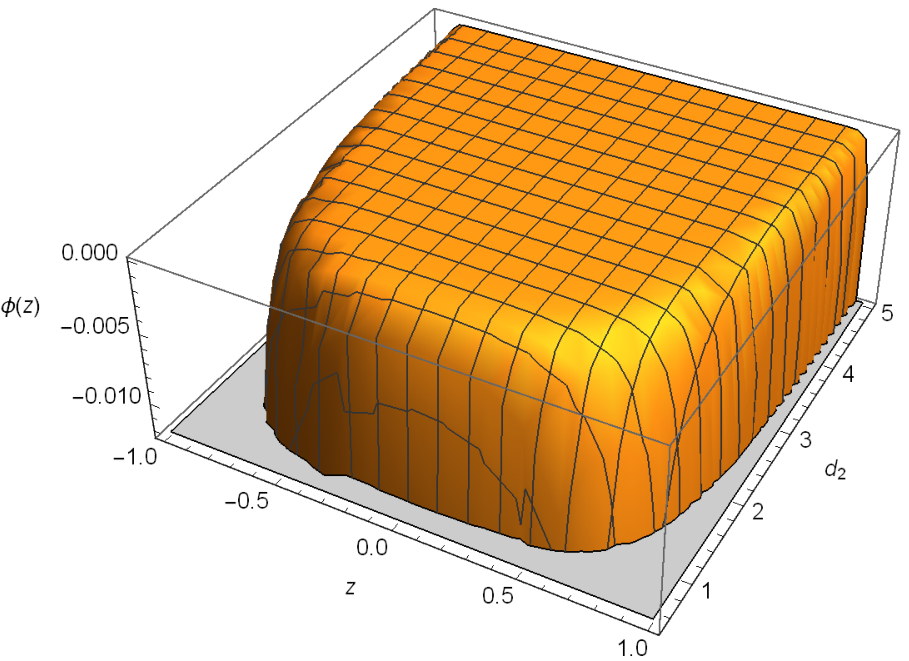}}
\\
   \centering
   \subfloat[The regular soliton for various $d_3$ values.]{\label{Fig1:c}
      \includegraphics[width=.47\textwidth]{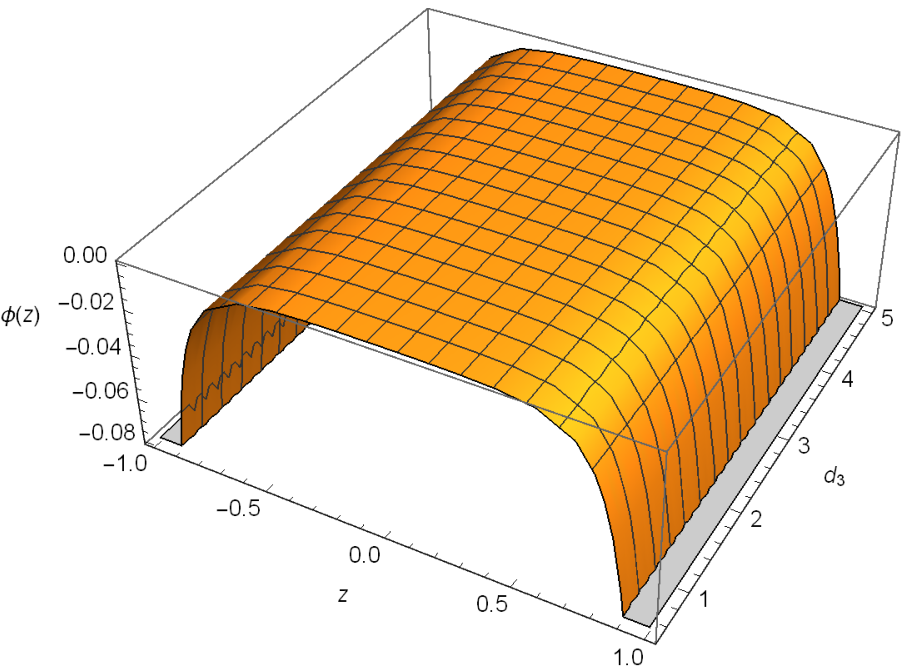}}
~
   \subfloat[The regular soliton for various $d_4$ values.]{\label{Fig1:d}
     \includegraphics[width=.47\textwidth]{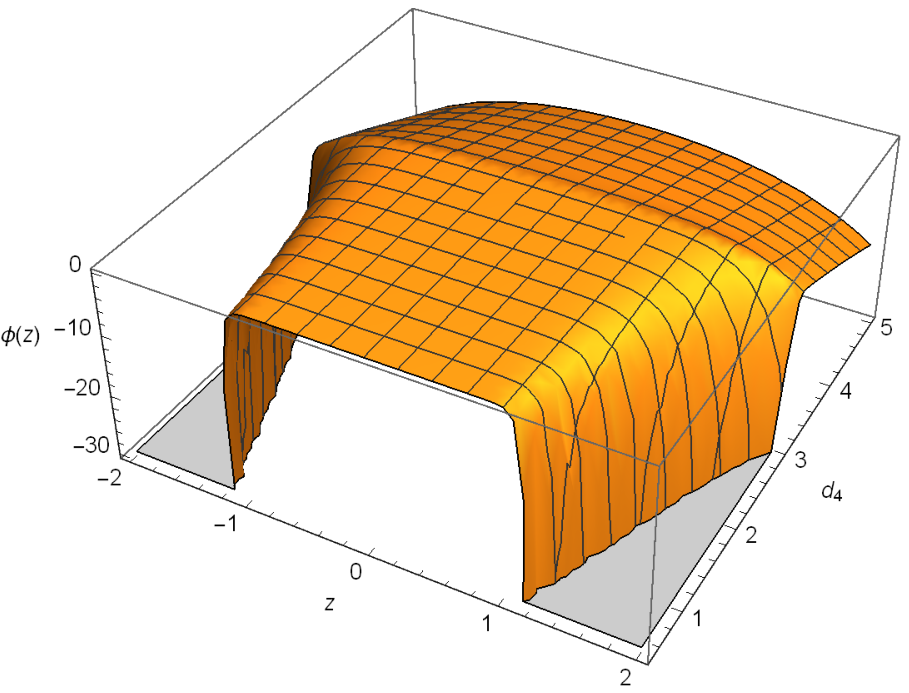}}
      
   \caption{Regular soliton for $z$ as one parameter is varied, with all others set to 1.}\label{Fig1}
\end{figure*}

In Figures \ref{Fig1}(a-d) the regular soliton for these values of $\rho$ and $A$ is presented, varying each of the parameters one by one from $0.5$ to $5$, and keeping the other parameters fixed at $1$. The range of $z$ is $-1$ to $1$ in Figures \ref{Fig1:a}--\ref{Fig1:c} and  $-2$ to $2$ in Figure \ref{Fig1:d}.

\subsection{Embedded solitons}
\label{subsec:4.2}
Now, following the same procedure and using the same trial function ansatz for embedded solitons as in I, we obtain the averaged action
\begin{multline}
	\dfrac{\sqrt{\pi}A\rho}{5184} \left( \dfrac{2592\sqrt{2}\alpha^2d_2}{d_1^{3/2}} +\dfrac{512\sqrt{3}A^2\left( 21d_2\rho^2-16d_4 \right)}{\rho^8} +\dfrac{1296\sqrt{2}A\left( 2\rho^4-15d_1 \right)}{\rho^6} \right. \\ +\dfrac{81A\alpha e^{-\dfrac{\rho^2}{4\sqrt{2}\sqrt{d_1}}}}{d_1^2 \rho^6}\left( 240\sqrt{2}d_1^2d_2-3\sqrt{2}d_4\rho^6-24d1^{3/2}\left( 3d_2\rho^2+20d_4 \right) \right. \\ \left. \left. +6\sqrt{2}d_1\rho^2\left( 17d_2\rho^2-4d_4 \right) +5\sqrt{d_1}\left( 3d_2\rho^6+4d_4\rho^4 \right) \right)  \right)=S,
\end{multline}

when
\begin{equation}
\kappa(c)=\frac{\sqrt[4]{2}}{\sqrt[4]{d_1}}.
\end{equation}

And so we can get non--trivial solutions for $A$ and $\rho^2$ that allow us to evaluate the embedded soliton for different values of the parameters.

\begin{figure*}[ht]
   \subfloat[The embedded soliton for various $d_1$ values.]{\label{Fig2:a}
      \includegraphics[width=.47\textwidth]{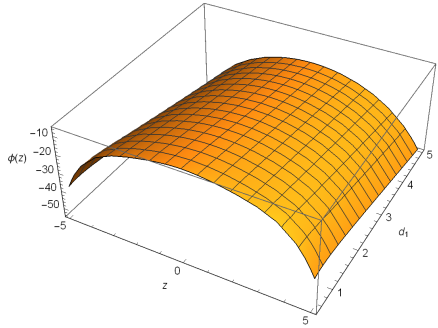}}
~
   \subfloat[The embedded soliton for various $d_2$ values.]{\label{Fig2:b}
     \includegraphics[width=.47\textwidth]{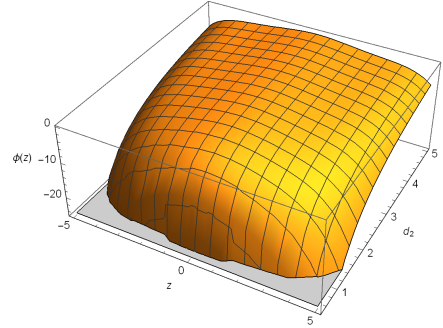}}
\\
   \centering
   \subfloat[The embedded soliton for various $d_3$ values.]{\label{Fig2:c}
      \includegraphics[width=.47\textwidth]{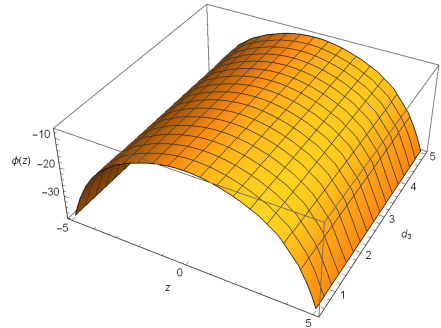}}
~
   \subfloat[The embedded soliton for various $d_4$ values.]{\label{Fig2:d}
     \includegraphics[width=.47\textwidth]{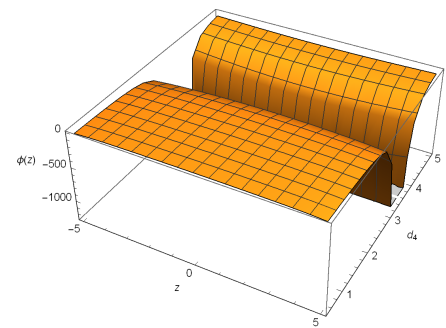}}
      
   \caption{Embedded soliton for $z$ with $\alpha=0.5$ as one parameter is varied, and all others set to 1.}\label{Fig2}
\end{figure*}

\begin{figure*}[ht]
	\subfloat[The embedded soliton for $d_1=3$.]{\label{Fig3:a}
	   \includegraphics[width=.47\textwidth]{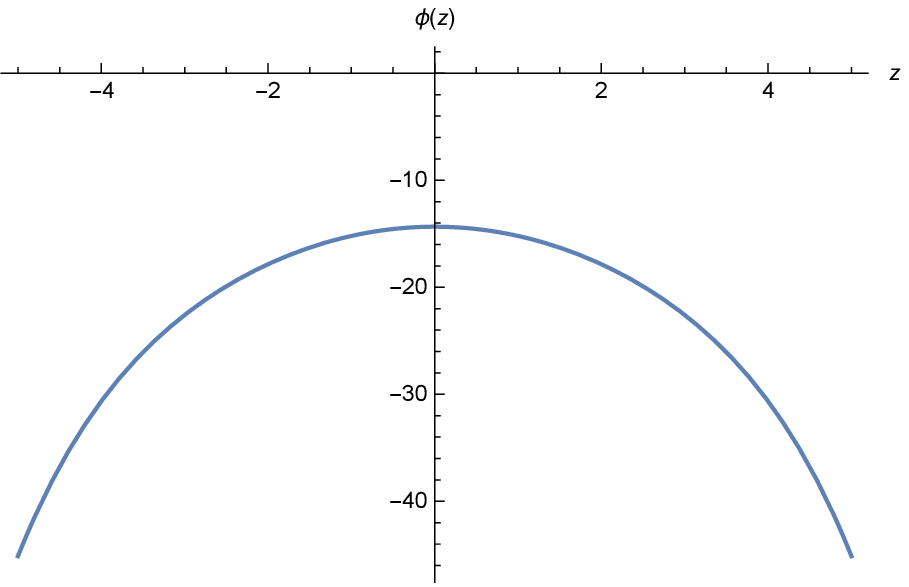}}
 ~
	\subfloat[The embedded soliton for $d_2=3$.]{\label{Fig3:b}
	  \includegraphics[width=.47\textwidth]{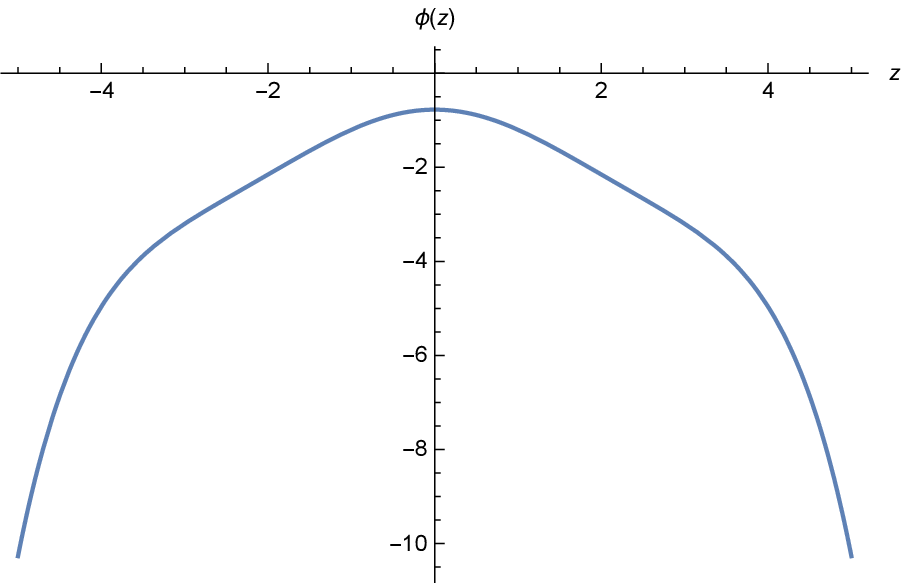}}
	   
	\caption{Embedded soliton for $z$, with $\alpha=0.5$, $d_1=3$ (a) or $d_2=3$ (b), and all others set to 1.}\label{Fig3}
 \end{figure*}

\begin{figure}[ht]
	\centering
	\includegraphics[width=.5\textwidth]{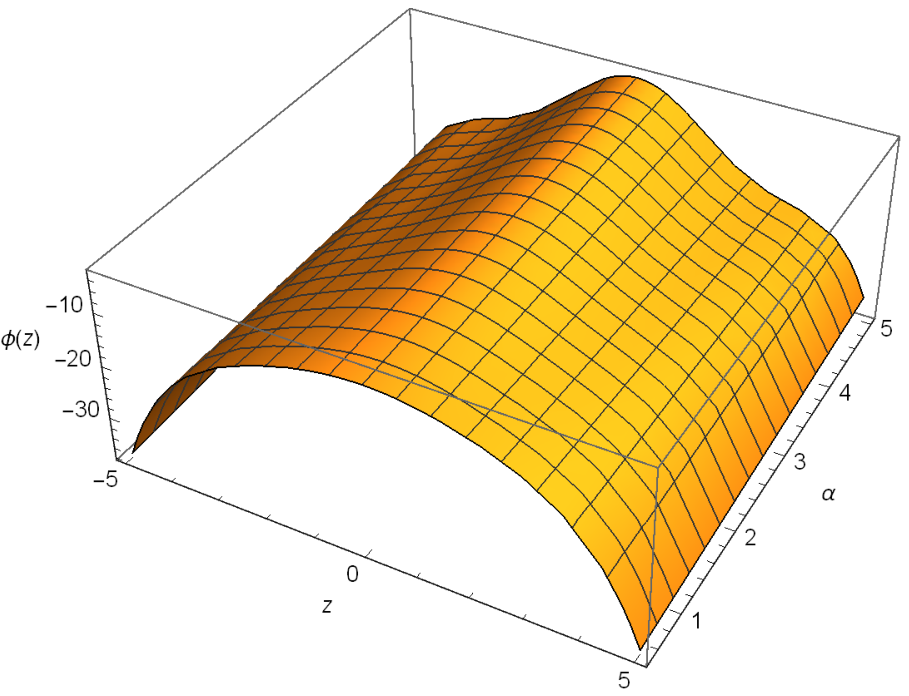}
	\caption{The embedded soliton as $\alpha$ varies, with all the other parameters set to 1.}\label{Fig4}
\end{figure}

In Figures \ref{Fig2}, \ref{Fig3} and \ref{Fig4} the embedded soliton is presented, with $z$ varying from -5 to 5. In Figure \ref{Fig2}(\ref{sub@Fig2:a}--\ref{sub@Fig2:d}), $\alpha=0.5$ and parameters are varied one by one from $0.5$ to $5$, and keeping the other parameters fixed at $1$. 
 We include Figures \ref{Fig3:a}--\ref{Fig3:b}, which are cross sections of Figures \ref{Fig2:a}--\ref{Fig2:b}, to facilitate the visualization of the results presented here. In Figure \ref{Fig4}, $d_1=d_2=d_3=d_4=1$, while $\alpha$ varies from $0.5$ to $5$.

\section{Generalizations of previous existence conditions for Lagrangians}
\label{sec:4}

\subsection{Third-order Lagrangians}

As mentioned in Section 1, the inverse problem of the Calculus of Variations for a sixth-order ODE was treated by Juras \cite{Jur}. We now proceed to compare our generalized family of Lagrangians \eqref{Eq:Coeff3}-\eqref{Eq:IntCond} for variational sixth-order ODEs against the criteria obtained by Juras using differential geometric approaches, as well as to various families of variational ODEs based on Juras' criteria \cite{CGK, GC}. 

To this end, we state Juras' principal result here:

\noindent
{\it Theorem: A sixth-order ODE 
	
	\begin{equation} \label{Juras6}
	u^{(6)} = F(z, u, u', u'', u''', u^{(4)}, u^{(5)}),
	\end{equation}
	admits a variational representation with a non-degenerate third order Lagrangian if and only if the following two conditions are satisfied
	
	\begin{equation} \label{Fels1}
	I_1 = 0 ,
	\end{equation}
	
	and
	
	\begin{equation} \label{Fels2}
	I_2 = 0,
	\end{equation}
 where $I_1$ and $I_2$ are given in Appendix B.
}

Converting our general variational equation \eqref{vareq} to the form in \eqref{Juras6}, we may read off the corresponding $F(z, u, u', u'', u''', u^{(4)}, u^{(5)})$ function.

For our generalized family of Lagrangians (which we copy here for ease of understanding of the following discussion) 

\begin{equation}\label{Eq:Coeff3_5}
L=c_1(u,u',u'')c_4(u^{(3)})+c_3(u,u',u'')u^{(3)}+c_5(u,u',u''),
\end{equation}
where 
\begin{equation}\label{Eq:IntCond_5}
c_4(u^{(3)})={\displaystyle \iint c_2(u^{(3)})du^{(3)}du^{(3)} }  .
\end{equation}

\noindent
It is now straightforward to verify, after somewhat lengthy computer algebra that neither the first condition \eqref{Fels1} nor the second condition \eqref{Fels2} above are satisfied for arbitrary functions $c_i$ in the Lagrangian \eqref{Eq:Coeff3_5}.

However, both conditions are satisfied for the special case
\begin{equation}
c_1(u,u', u'') = 1, c_2(u^{(3)}) = 1,
\end{equation}
considered in Juras' \cite{Jur} differential geometric derivation of this criterion, as well as in the treatments of various models based his  conditions\cite{CGK}.

{\it This leads us to the following conclusions about the two ways that our generalized family of Lagrangians \eqref{Eq:Coeff3_5}-\eqref{Eq:IntCond_5} and the associated variational ODEs \eqref{vareq} are more general than the criteria developed in Juras\cite{Jur}:\\
	
	\noindent
	a. our Lagrangian has four arbitrary or free functions $c_1, c_3, c_4$, and $c_5$, in place of Juras' single function $F$;
	
	\noindent
	and
	
	\noindent
	b. the leading coefficient $c_1(u,u',u'') c_4(u^{(3)})$ in our variational ODE \eqref{vareq} may be more general than for Juras or other models based on his criteria for a variational
	representation.
}

\subsection{Fourth-order Lagrangians}

We next briefly discuss Juras' analogous differential geometric criterion for eighth-order ODEs:

\noindent
{\it Theorem: An eighth-order ODE 
	
	\begin{equation} \label{Juras8}
	u^{(8)} = F(z, u, u', u'', u''', u^{(4)}, u^{(5)}, u^{(6)}, u^{(7)}),
	\end{equation}
	admits a variational representation with a non-degenerate third order Lagrangian if and only if the five conditions are satisfied
	
	\begin{equation} \label{Fels3}
	I_1 = I_2 = I_3 = J_1 = J_2 =0 ,
	\end{equation}
	
\noindent	
where $I_1$ through $J_2$ are detailed in \cite{Jur}.
}

In an entirely analogous way to the previous subsection, it is straightforward to check by lengthy computer algebra that the above criteria are only satisfied for the special cases

\begin{equation}
d_1(u,u', u'', u^{(3)}) = 1, d_2(u^{(4)}) = 1.
\end{equation}


\section{Conclusions}
\label{sec:conclusions}

In conclusion, we have extended the techniques developed in I to derive generalized hierarchies of Lagrangians for sixth- and eighth-order nonlinear ODEs. A representative member of the family has also been employed to construct families of regular and embedded solitary waves of the corresponding variational equation.

In particular, our families of Lagrangians and the associated variational ODEs are more general than the most general variational formulations derived earlier for sixth- and eighth-order ODEs \cite{Jur} in two significant ways:

\noindent
a. our Lagrangians have four arbitrary or free functions $c_1, c_3, c_4$, and $c_5$, or $d_1, d_3, d_4$, and $d_5$  in place of the  single function $F$ in the earlier general variational criteria  

\noindent
and

\noindent
b. the leading coefficients $c_1(u,u',u''), c_4(u^{(3)})$ or $d_1(u,u',u'',u^{(3)}), d_4(u^{(4)})$ in our variational ODEs \eqref{vareq} may be more general than in earlier work, or in other models based on the earlier general criteria for a variational representation\cite{CGK}.




\appendix

\section{Full Equations}
\label{sec:AppA}

The complete Euler--Lagrange Equation for a nonlinear ODE of up to sixth order reads
\begin{multline}\label{Eq:EL}
	-u^{(6)}L_{u^{(3)}u^{(3)}} -{u^{(4)}}^3L_{u^{(3)}u^{(3)}u^{(3)}u^{(3)}} -3u^{(3)}u^{(5)}L_{u''u^{(3)}u^{(3)}} +{u^{(3)}}^2L_{u''u''u''} -{u^{(3)}}^3L_{u''u''u''u^{(3)}}\\
	-3u''u^{(5)}L_{u'u^{(3)}u^{(3)}} -3{u^{(3)}}^2L_{u'u''u^{(3)}} +2u'u^{(3)}L_{u'u''u''} -3u''{u^{(3)}}^2L_{u'u''u^{(3)}} -u''L_{u'u'}\\ 
	-3u''u^{(3)}L_{u'u'u^{(3)}} +{u''}^2L_{u'u'u''} -3{u''}^2u^{(3)}L_{u'u'u''u^{(3)}} -{u''}^3L_{u'u'u'u^{(3)}}-u^{(3)}L_{uu^{(3)}} \\
	-{u^{(4)}}^2\left( 2L_{u''u^{(3)}u^{(3)}} +3u^{(3)}L_{u''u^{(3)}u^{(3)}u^{(3)}} +3u''L_{u'u^{(3)}u^{(3)}u^{(3)}} +3u'L_{uu^{(3)}u^{(3)}u^{(3)}} \right) \\
	-3u'u^{(5)}L_{u'u^{(3)}u^{(3)}} +u''L_{u,u''} -3u''u^{(3)}L_{uu''u^{(3)}} +2u'u^{(3)}L_{uu''u''} -u'L_{uu'}\\
	-3u'{u^{(3)}}^2L_{uu''u''u^{(3)}} -3{u''}^2L_{uu'u^{(3)}} -3u'u^{(3)}L_{uu'u^{(3)}} +2u'u''L_{uu'u''} -6u'u''u^{(3)}L_{uu'u''u^{(3)}} \\
	-3u'{u''}^2L_{uu'u'u^{(3)}} -3u'u''L_{uuu^{(3)}} -u^{(4)}\left( 3u^{(5)}L_{u^{(3)}u^{(3)}u^{(3)}} +3{u^{(3)}}^2L_{u''u''u^{(3)}u^{(3)}} \right. \\ 
	\left. +u^{(3)}L_{u''u''u^{(3)}} +3u^{(3)}L_{u'u^{(3)}u^{(3)}} +u''L_{u'u''u^{(3)}} +6u''u^{(3)}L_{u'u''u^{(3)}u^{(3)}} \right.\\
	\left. +3{u''}^2L_{u'u'u^{(3)}u^{(3)}} +3u''L_{uu^{(3)}u^{(3)}} +u'L_{uu''u^{(3)}} +6u'u^{(3)}L_{uu''u^{(3)}u^{(3)}} +6u'u''L_{uu'u^{(3)}u^{(3)}} \right.\\
	\left. +3{u'}^2L_{uuu^{(3)}u^{(3)}} +2L_{u'u^{(3)}} -L_{u''u''} \right) +{u'}^2L_{uuu''} -3{u'}^2u^{(3)}L_{uuu''u^{(3)}} \\
	-3{u'}^2u''L_{uuu'u^{(3)}} -{u'}^3L_{uuuu^{(3)}} +L_{u}=0.
\end{multline}

The terms accompanying the different orders of the derivatives in Equation \eqref{vareq} are
\begin{subequations}
\begin{align}
    Z_1= & L_{u^{(3)}u^{(3)}}, \\
    E_1= & -3\left( u^{(4)}L_{u^{(3)}u^{(3)}u^{(3)}} +u^{(3)}L_{u''u^{(3)}u^{(3)}} +u''L_{u'u^{(3)}u^{(3)}} +u'L_{uu^{(3)}u^{(3)}}\right), \\
    \Delta_3= & -L_{u^{(3)}u^{(3)}u^{(3)}u^{(3)}}, \\
    \Delta_2= & -2L_{u''u^{(3)}u^{(3)}} -3\left( u^{(3)}L_{u''u^{(3)}u^{(3)}u^{(3)}} +u''L_{u'u^{(3)}u^{(3)}u^{(3)}} +u'L_{uu^{(3)}u^{(3)}u^{(3)}} \right), \\
    \Delta_1= & -L_{u''u''}-2L_{u'u^{(3)}} -3{u^{(3)}}^2L_{u''u''u^{(3)}u^{(3)}} \notag \\
     & \quad -u^{(3)}\left( 6u''L_{u'u''u^{(3)}u^{(3)}} +6u'L_{uu''u^{(3)}u^{(3)}} +3L_{u'u^{(3)}u^{(3)}} +L_{u''u''u^{(3)}} \right) \notag \\
     & \qquad -3{u''}^2L_{u'u'u^{(3)}u^{(3)}} -u''\left( 6u'L_{uu'u^{(3)}u^{(3)}} +3L_{uu^{(3)}u^{(3)}} +L_{u'u''u^{(3)}} \right) \notag \\
     & \quad \qquad -3{u'}^2L_{uuu^{(3)}u^{(3)}} -u'L_{uu''u^{(3)}} , \\
    \Gamma_3= & -L_{u''u''u''u^{(3)}}, \\
    \Gamma_2= & L_{u''u''u''} -3\left( u''L_{u'u''u''u^{(3)}} +u'L_{uu''u''u^{(3)}} +L_{u'u''u^{(3)}} \right), \\
    \Gamma_1= & -L_{uu^{(3)}} -3{u''}^2L_{u'u'u''u^{(3)}} \notag \\
     & \quad -u''\left( 6u'L_{uu'u''u^{(3)}} +3L_{uu''u^{(3)}} +3L_{u'u'u^{(3)}} -2L_{u'u''u''} \right) \notag \\
     & \quad \qquad -3{u'}^2L_{uuu''u^{(3)}} +u'\left( 2L_{uu''u''} -3L_{uu'u^{(3)}} \right) , \\
    B_3= & -L_{u'u'u'u^{(3)}}, \\
    B_2= & L_{u'u'u''} -3\left( u'L_{uu'u'u^{(3)}} +L_{uu'u^{(3)}} \right), \\
    B_1= & L_{uu''}-L_{u'u'} -3{u'}^2L_{uuu'u^{(3)}} +u'\left( 2L_{uu'u''} -3L_{uuu^{(3)}} \right) , \\
    A_3= & -L_{uuuu^{(3)}}, \\
    A_2= & L_{uuu''}, \\
    A_1= & -L_{uu'}.
\end{align}
\end{subequations}

\section{Juras's Differential Geometric Criteria}
\label{sec:AppB}

The two quantities $I_1$ and $I_2$ in equations \eqref{Fels1} and \eqref{Fels2} for Juras's criteria for a variational representation of a sixth-order ODE are

\begin{align} \label{Fels1Def}
   I_1 = 
   & - 2 F_{u^{(5)}zzzz}/3 + 10 F_{u^(5)} F_{u^{5}zzz}/ 9 + F_{u^{(4)}zzz} + 20  F_{u^{(5)}z}  F_{u^{(4)}zz} / 9 \nonumber \\
   & - 20  (F_{u^{(5)}})^2  F_{u^{(5)}zz} / 27 -  F_{u^{(4)}}  F_{u^{(5)}zz} / 3 -  F_{u^{(5)}}  F_{u^{(4)}zz} -  F_{u^{(3)}zz} \nonumber\\
   & - 10  F_{u^{(5)}}  (F_{u^{(5)}z})^2 / 9 -  F_{u^{(5)}z}  F_{u^{(4)}z} + 20  (F_{u^{(5)}})^3  F_{u^{(5)}z} / 81 \nonumber\\
   & +  (F_{u^{(5)}})^2  F_{u^{(4)}z} / 3 +  F_{u^{(5)}}  F_{u^{(4)}} F_{u^{(5)}z} / 3 +  F_{u^{(3)}}  F_{u^{(5)}z} / 3 + 2  F_{u^{(5)}}  F_{u^{(3)}z} / 3 +  F_{u^{(2)}z} \nonumber\\
  & - 2  (F_{u^{(5)}})^5 / 243  -  (F_{u^{(5)}})^3  F_{u^{(4)}} / 27 -  (F_{u^{(5)}})^2  F_{u^{(3)}} / 9 -  F_{u^{(5)}}  F_{u^{(2)}} / 3 -  F_{u^{(1)}} ,
\end{align} 
and
\begin{align} \label{Fels2Def}
I_2  = 
& 5 F_{u^{(5)}zz}/3 - 5 F_{u^(5)} F_{u^{5}z}/ 3 - 2 F_{u^{(4)}} \nonumber \\
&  + 5  (F_{u^{(5)}})^3 / 27 + 2  F_{u^{(5)}}  F_{u^{(4)}} / 3 + F_{u^{(3)}}.
\end{align}



\section*{References}
\begin{thebibliography}{99}
%
%
	
	
	
	
	\bibitem{PUM} Manheim, P and Davidson, A, Dirac quantization of the Pais-Uhlenbeck fourth order oscillator, Phys Rev A 71 (2005) 042110; 
	
	\bibitem{Man} Manheim, P, Solution to the ghost problem in fourth order derivative theories, Foundations Phys 37 (2007) 532.
	
	\bibitem{TCG} Tanriver, U, Roy Choudhury, S and Gambino, G, Lagrangian dynamics and possible isochronous behavior in several classes of non-linear second order oscillators via the use of the Jacobi last multiplier, Intl, J. Nonlin. Mech. 74 (2015) 100.
	
	\bibitem{CGK} Ghosh Choudhury, A, Guha, P and Kudryashov, N, A Lagrangian description of the higher-order Painlev\'e equations, Appl. Math. Comput. 218 (2012) 6612.
	
	\bibitem{VK} Vogel, T and Kaup, D J, Quantitative measurement of variational approximations, Physics Letters A 362 (2007) 289.
	
	\bibitem{Helm} Helmholtz, H, Ueber die physikalische bedeutung des prinicips der kleinsten wirkung, J. Reine Angew Math, 100 (1887) 137.
	
	\bibitem{Lop} Lopuszanski, J, The inverse variational problem in classical mechanics (World Scientific, Singapore, 1999).
	
	\bibitem{Dar} Darboux, G, Lecon sur la theorie generale des surfaces (Gauthier-Villars, Paris, 1894).
	
	\bibitem{Jur} Juras, M, The inverse problem of the calculus of variations for sixth- and eighth-order scalar ordinary differential equations, Acta Applicandae Math., 66 (2001) 25.
	
	\bibitem{Ran} Alfonso-Rodriguez, R, and Roy Choudhury, S, Novel Lagrangian Hierarchies, Generalized Variational ODE's, and Families of Regular and Embedded Solitary Waves, J. Phys. A: Math. and Theor., 53 (2020) 375701
	
	\bibitem{Fels} Fels, M, The inverse problem of the calculus of variations for scalar fourth-order ordinary differential equations, Trans. Amer. Math Soc, 348 (1996) 5007.

	
	
	
	

	\bibitem{VPh} Vogel, T, Soliton Solutions Of Nonlinear Partial Differential Equations Using Variational Approximations And Inverse Scattering Techniques, PhD Thesis, University of Central Florida (2007)
	
	\bibitem{Smi09} Smith, T B, and Roy Choudhury, S, Regular and embedded solitons in a generalized Pochammer PDE, Commun. Nonlinear Sci. Numer. Simul, 14 (2009) 2637.
	
	

	
   



	
	\bibitem{GC} Guha, P and Ghosh Choudhury, A, On Lagrangians and Hamiltonians of some fourth-order nonlinear Kudryashov ODEs, Communications in Nonlinear Science and Numerical Simulation, 16, (2011) 3914.
	
\end{thebibliography}
\end{document}